\documentclass{elsart3}

\usepackage{mathptmx}
\usepackage{bm}
\usepackage{graphicx}
\usepackage{amssymb}

\begin{document}

\begin{frontmatter}



\title{Crystal-field-induced magnetostrictions in the spin reorientation process of 
Nd$_2$Fe$_{14}$B-type compounds}


\author[Sendai]{H. Kato\corauthref{cor1}},
\ead{kato@mlab.apph.tohoku.ac.jp}
\author[Tagajo]{M. Yamada},
\author[Sendai] {T. Miyazaki}
\address[Sendai]{Department of Applied Physics, Tohoku University,
       Aoba-yama 05, Sendai 980-8579, Japan}
\address[Tagajo]{Faculty of Engineering, Tohoku Gakuin University,
        Tagajo 985-0873, Japan}       
 \corauth[cor1]{Corresponding author.  Fax: +81-22-217-7947.}

\begin{abstract}

Volume expansion $\Delta V / V$ associated with the spin reorientation process of Nd$_2$Fe$_{14}$B-type compounds has been investigated in terms of simple 
crystalline-electric-field (CEF) model.  In this system, $\Delta V / V$ is shown to be a direct measure of second order CEF energy.   Calculated anomalies in $\Delta V / V$ associated with the first-order magnetization process of Nd$_2$Fe$_{14}$B are presented, which well reproduced the observations.
\end{abstract}

\end{frontmatter}

Magnetoelastic (ME) effects in 3$d$-4$f$ intermetallic system have been investigated extensively during the last three decades.
Kuz'min \cite{Kuzmin}  has proposed a linear theory of  magnetostriction 
by assuming that CEF and ME interactions are small with respect to the 3$d$-4$f$ exchange interaction.
We studied the effect of CEF on the magnetostriction in melt-spun ribbons of (Er$_{1-x}$Tb$_x$)$_2$Fe$_{14}$B with $0 \le x \le 0.15$.  
We have found that the sum of the longitudinal and transverse magnetostriction constants $\lambda_{//} +\lambda_{\perp}$ took a maximum at successive spin reorientation (SR) temperatures \cite{Kato2001}.  A simple model calculation of the volume change $\Delta V / V$ was made by assuming that only the lowest order CEF potential terms $A_2^0$ for Er and Tb sites depend on $\Delta V / V$ \cite{Kato2001}.  This calculation showed that $\Delta V / V$ exhibits a sharp peak at SR temperatures, which is in accordance with the experimental peaks of  $\lambda_{//} +\lambda_{\perp}$.  In this paper, we propose that this
simple model, in which a Hamiltonian containing
CEF, ME, 3$d$-4$f$ exchange, and Zeeman interactions is diagonalized numerically
 without a perturbation framework, can be applicable to other Nd$_2$Fe$_{14}$B-type of intermetallics.
 
We showed that calculated SR temperatures are almost independent of ME effect \cite{Kato2001},
since  $\Delta V / V$ is small.  In such circumstances,  volume dependent terms can be extracted from the total free energy expression in $R_2$Fe$_{14}$B system: 

\begin{equation}
\Delta E = -B_2^0 \langle O_2^0\rangle  \Delta V/V + C (\Delta V/V)^2, \label{eq:delta-E}
\end{equation}

\noindent
where the first term expresses a CEF-induced ME energy with
$B_2^0$ and $O_2^0$,  denoting the second-order CEF coefficient and Stevens operator, respectivly.   
The second term of eq. (\ref{eq:delta-E}) is the elastic energy with the elastic constant $C$.
On condition of the minimization of 
eq. (\ref{eq:delta-E}), we obtain that $\Delta V/V=B_2^0 \langle O_2^0\rangle /2C.$
This states that the volume change is proportional to the second order CEF energy of the $R$ sublattice. In the case of (Er$_{1-x}$Tb$_x$)$_2$Fe$_{14}$B system, the fact that $B_2^0{\rm (Er)} > 0, B_2^0{\rm (Tb)} < 0$, and different temperature dependence of
$\langle O_2^0{\rm (Er)}\rangle $ and $\langle O_2^0{\rm (Tb)}\rangle $ terms give rise to a distinguished variation of $\Delta V/V$, especilally around the SR transitions.

The discussion above is, in general, applicable to a series of Nd$_2$Fe$_{14}$B-type
intermetallics, in which observed volume magnetostriction is less than $10^{-3}$ \cite{Algarabel1992}.  It is therefore possible to deduce an information about the CEF interaction by measuring the $\Delta V/V$ in these system.  The most effective
way to separate CEF-induced volume change from observed volume magnetostriction
 is to measure the anomalous part of $\Delta V/V$ around the SR transition, which
is induced either by temperature or by magnetic field.  We have numerically calculated the
field dependence of $\Delta V/V$ associated with the first-order magnetization process (FOMP) in Nd$_2$Fe$_{14}$B and Pr$_2$Fe$_{14}$B.  The CEF, molecular-field and elastic
parameters used in the calculations are the same as those reported previously
\cite{Kato2001,Yamada1988}.   Calculated field dependence of Fe-moment direction $\theta$ and $\Delta V/V$ is given in Fig. 1 for several temperatures. 
\begin{figure}[t]
\begin{center}
\includegraphics*[scale=.68]{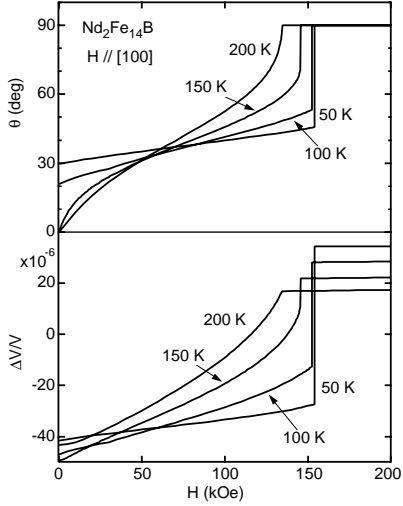}
\end{center}
\caption{Calculated field dependence of $\theta$, the angle between the Fe moment and the [001] directions, and
the volume change $\Delta V / V$ in Nd$_2$Fe$_{14}$B.  The field is applied
along the [100] direction.}
\label{NdFeB}
\end{figure}%
An abrupt increase of $\theta$
and $\Delta V/V$ is seen at the FOMP, which becomes smeared at elevated 
temperatures.  This behavior is consistent with the observed high-field magnetization curves \cite{Kajiwara}.  Algarabel {\it et al.} \cite{Algarabel1990} reported magnetostriction isotherms for Nd$_2$Fe$_{14}$B single crystals, which exhibit
clear anomalies associated with the FOMP.  
Although the observed expansion is anisotropic at the FOMP with the $a$-axis elongation 
 and $c$-axis shrinkage,  the volume anomaly deduced from the relation
 $\Delta V/V=\lambda(\bm{a},\bm{c})+2\lambda(\bm{a},\bm{a})$
  is about $20\times 10^{-6}$ at 150 K, which is to compare with the  $\Delta V / V$ jump
 in Fig. 1 ($11\times 10^{-6}$ at 150 K).
 In view of the fact that no additional parameters are required,  
 the present calculation is in good agreement with the experiment.
We have thus established a convenient method to extract CEF information in
rare-earth intermetallics.  To demonstrate this, we show in Fig. 2
the predicted $\Delta V / V$ anomalies at the FOMP in Pr$_2$Fe$_{14}$B
 with fields applied both along the [100] and [110] directions.
According to these results Pr$_2$Fe$_{14}$B exhibits a much larger
discontinuity in $\Delta V / V$ at the FOMP than Nd$_2$Fe$_{14}$B.

\begin{figure}[t]
\begin{center}
\includegraphics*[scale=.68]{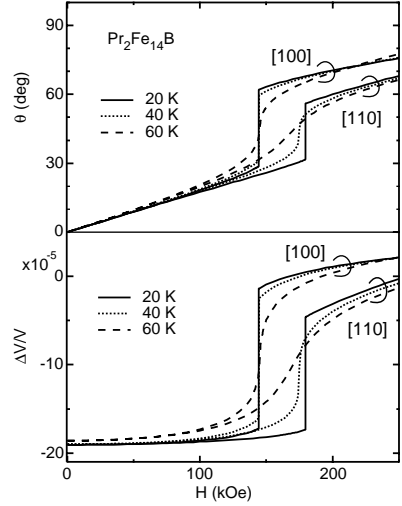}
\end{center}
\caption{Similar calculations to Fig. 1 for Pr$_2$Fe$_{14}$B.}
\label{PrFeB}
\end{figure}%



\end{document}